\documentclass[twocolumn,showpacs,preprintnumbers,amsmath,amssymb]{revtex4}
\usepackage{epsfig,amsfonts}
\usepackage{amsmath}
\usepackage{bbm}
\usepackage{wasysym}
\usepackage{graphicx,psfrag,rotating}
\usepackage{graphics}
\usepackage{txfonts}

\usepackage{textcomp}
\usepackage{graphicx}
\usepackage{dcolumn}
\usepackage{bm}

\begin{document}


\title{Interaction of double sine-Gordon solitons with external potentials: an analytical model}

\author{Samira Nazifkar, Kurosh Javidan and Mohsen Sarbishaei}
\affiliation{ Physics Department, School of Science, Ferdowsi University of Mashhad, Mashhad, Iran }

\date{\today}

\begin{abstract}

Interaction of Double sine-Gordon solitons with a space dependent  potential wall and also a potential well has been investigated by employing an analytical model based on the collective coordinate approach. The potential has been added to the model through a suitable nontrivial metric for the background space-time. The model is able to predict most of the features of the soliton-potential interaction. It is shown that a soliton can pass through a potential barrier if its velocity is greater than a critical velocity which is a function of soliton initial conditions and also characters of the potential. It is interesting that the solitons of the double sine-Gordon model can be trapped by a potential barrier and oscillate there. This situation is very important in applied physics. Soliton-well system has been investigated using the presented model too. Analytical results also have been compared with the results of the direct numerical solutions. 

\end{abstract}
\maketitle

\section{\label{intro}Introduction }
The sine-Gordon equation attracted much interest of physicists. This equation is a non-linear partial differential equation which appears naturally in different physical systems: in atomic physics\cite{1}, electromagnetism\cite{2}, superconductivity\cite{3}, field theory\cite{4}, biophysics\cite{5,6,7} and statistical mechanics\cite{8}. It also has plenty of application in condensed matter systems \cite{9,10} and nonlinear optics\cite{11}. Moreover the solitons and kinks of the $\mathrm{SU(3)}$ generalized sine-Gordon model (GSG) are shown to describe the baryonic spectrum of two-dimensional quantum chromodynamics (QCD2) \cite{12}. 

As a natural development of the studies on integrable quantum field theories, there has been recently an increasing interest in studying the properties of such non-integrable quantum field theories in (1 + 1) dimensions (like double sine-Gordon (DSG) model) both for theoretical reasons and their applications. The Lagrangian of a realistic physical system often gives a more complicated equation of motion than the sine-Gordon equation. For example, a quantum spin chain is mapped into a Lagrangian with several potential terms\cite{13}. Systems with nonlinear optical properties also give rise to more complicated wave equations\cite{14}. Thus a more enhanced model is desirable. This leads to the DSG equation. 

The DSG model has been applied to model a variety of systems in condensed matter, quantum optics, and particle physics\cite{15}. Condensed-matter applications include the spin dynamics of superfluid 3He\cite{16}, magnetic chains\cite{17}, commensurate-incommensurate phase transitions\cite{18}, surface structural reconstructions\cite{19}, domain walls\cite{20,21} and fluxion dynamics in Josephson junction\cite{22}. In quantum field theory and quantum optics DSG applications include quark confinement\cite{23} and self-induced transparency\cite{24}.

In ideal DSG equation the parameters of the model are space-time independent fixed parameters. But it is clear that in a realistic system, such parameters are functions of space or time. For example consider the long Josephson junction. For a sufficiently wide class of Josephson junctions the superconducting Josephson current (phase difference of superconductor's wave functions) can be represented as a sine series. Using only first two terms of this expansion one can show\cite{23} that the distribution of the magnetic flux along x-axis of the junction in the static regime satisfies the DSG equation which their parameters are depend on the preparation technology of junctions which naturally cannot be fixed along the junction. This means that the parameters of the model become space dependent because of medium disorders. In this situation the localized solution encounters some kinds of space dependent potentials which greatly affects on the behavior of the soliton. 

There has been an increasing interest in the scattering of solitons from defects or impurities, which generally come from medium properties. As menntioned before, the motivations come from both theoretical and applied aspects of physics. The effects of medium disorders and impurities can be added to the equation of motion as perturbative terms\cite{25,26}. These effects can also be generated by making some parameters of the equation of motion to a function of space or time\cite{27,28}. There still exists another interesting method which is mainly suitable for working with topological solitons\cite{29,30}. In this method, one can add such effects to the Lagrangian of the system by introducing a suitable nontrivial metric for the background space-time without losing the topological boundary conditions. 

Numerical simulation is the main tool which is used for investigation of soliton behaviour in a defective medium . Although one could always rely on numerical methods to shed some light on their properties, it would be obviously useful to develop some theoretical tools to control them analytically. Motivated by this situation an analytical model is presented to investigate the interaction of solitons of DSG model with defects using a collective coordinate approach. Also the results of analytical model have been compared with the results of direct numerical solution of the real model. 

\section{\label{first}The double sine-Gordon model}
General form of the Lagrangian density for a real scalar field $\phi$ is 
\begin{equation}\label{1} 
\mathcal{L}=\frac{1}{2}\partial_\mu \phi\partial^{\mu}\phi-U(\phi).
\end{equation}
The DSG potential which contains a constant and a harmonic term in addition to the self-interaction potential of the ordinary sine-Gordon equation is considered here\cite{31,32}
\begin{equation}\label{2}
U(\phi)=1-cos(\phi)+A\left (1-cos(2\phi)\right),
\end{equation}
where $A$ is a constant. This potential has absolute degenerate minima at $\phi= 2n\pi$ as the true vacuua, and the metastable, local minima at $\phi=(2n + 1)\pi$ as the false vacuua \cite{32}. The harmonic term in this potential can result from the Fourier expansion of an arbitrary, periodic potential $V(\phi)=V(\phi+2n\pi)$. One does not
expect the system to remain integrable by adding these extra terms \cite{32}. The potential reduces to the ordinary SG potential in the limit $ A \rightarrow 0$. We focus on a model with $A=1$.
One soliton solution for the DSG equation can be written as \cite{33, 34}
\begin{equation}\label{4}
\phi=k\pi-2\tan^{-1}\left(\frac{1}{\sqrt{5}}\sinh\left(\sqrt{5}\frac{x-\left(x_0-\dot{X}(t)\right)}{\sqrt{1-\dot{X}^2}}\right)\right),
\end{equation}
where $x_{0}$ and $\dot{X}$ are soliton initial position and its velocity respectively.

As mentioned before, one can add effects of medium disorders to the Lagrangian of the system by introducing a suitable nontrivial metric for the background space time.  In other words, the metric carries the information of the medium. The suitable metric in the presence of a weak potential $V(x)$ is\cite{29,30,35a}
\begin{equation}\label{5}
g_{\mu\nu}(x)=\left(
\begin{array}{cc}
1+V(x) & 0\\
0 & -1\\
\end{array}\right).
\end{equation}
It is the weak field approximation indeed in which $V(x)=2\phi(x)$ where $\phi (x)$ is the corresponding potential in the flat space-time. 
The general form of the action in an arbitrary metric is 
 \begin{equation}\label{6}
S=\int\mathcal{L}\sqrt{-g}\mathrm{d}^nx,
\end{equation}
where $g$ is the determinant of the metric $g_{\mu\nu}(x)$. Therefore, we have the
effective Lagrangian:
\begin{equation}\label{7}
\mathcal{L}_{\mathrm{eff}}=\sqrt{-g}\left(\frac{1}{2}\partial_\mu \phi\partial^{\mu}\phi-U(\phi)\right).
\end{equation}
The motion equation of the field $\phi$ from the Lagrangian (\ref{1}) is\cite{35,36,37}:
\begin{equation}\label{3}
\frac{1}{\sqrt{-g}}\left(\sqrt{-g}\partial_\mu\partial^{\mu}\phi+\partial_\mu\phi\partial^{\mu}\sqrt{-g}\phi\right)  +\frac{\partial U(\phi)}{\partial\phi}=0.
\end{equation} 
Energy density of the "field+potential" can be found by varying "both" the field and the metric \cite{36}. By inserting the solution (\ref{4}) in the effective Lagrangian (\ref{7}) with the DSG model (\ref{2}) and using the metric (\ref{5}), with adiabatic approximation {\cite{25, 26}, we have
\begin{equation}\label{8}
\mathcal{L}_{\mathrm{eff}}=\sqrt{g_{00}}\left(\frac{1}{2}\left(g^{00}\dot{\phi}^2-g^{11}{\phi^{\prime}}^2\right)-\left(2-\cos(\phi)-\cos\left(2\phi\right)\right)\right)
\end{equation}
and finally, for a slowly varying weak potential $V(x)$, the effective Lagrangian reduces to
\begin{equation}\label{9}
\mathcal{L}_{\mathrm{eff}}\approx\left(\dot{X}^2-2-\left(\frac{1}{2}\dot{X}^2+1\right)V(x)\right)\frac{50\cosh^2\left(\sqrt{5}\left(x-X\right)\right)}{\left(5+\sinh^2\left(\sqrt{5}(x-X)\right)\right)^2}.
\end{equation} 
\section{\label{second}Collective coordinate variable}
The derivation of the collective action for the motion of the vortex centers starts with the elegant idea of Manton\cite{38}. A collective action can be constructed by substituting the collective vortex ansatz for the field configuration with vortices at $X_\mathrm{i}(t), \mathrm{i}=1,\dots,N$, into the effective field theory action and reduce the action to a function of the collective coordinates, $L[X_\mathrm{i}(t)] =\int \mathcal{L}_{\mathrm{eff}}(\phi(x,t,X_\mathrm{i}(t))\mathrm{d}x$ \cite{39}. The soliton internal structure can be omitted by integrating the Lagrangian density (or Hamiltonian density) over the variable $x$. After integration, the center of the soliton can be considered a particle if we look at this as a collective coordinate variable. As a result, the model is able to give us an analytic description for the evolution of the soliton center during the soliton-potential interaction.

Most of the local space dependent external potentials are appeared as a Dirac delta-like function. Thus we can choose $V(x)=\varepsilon \delta(x)$. If $\varepsilon>0$, we have a barrier and $\varepsilon<0$ creates a potential well. In this situation, the equation of motion for the variable $X(t)$ is derived from (\ref{9}) as:
\begin{equation}\label{10}\nonumber
\ddot{X}\left(M_0-50\varepsilon\frac{\cosh^2\left(\sqrt{5}X\right)}{\left(5+\sinh^2\left(\sqrt{5}X\right)\right)^2}\right)=\\
\end{equation}
\begin{equation}
\left(\dot{X}^2-2\right)50\sqrt{5}\varepsilon\frac{\cosh\left(\sqrt{5}X\right)\sinh\left(\sqrt{5}X\right)\left(3-\sinh^2\left(\sqrt{5}X\right)\right)}{\left(5+\sinh^2\left(\sqrt{5}X\right)\right)^3}.
\end{equation}
Where $M_0=\int_{-\infty}^\infty\frac{100\cosh\left(\sqrt{5}(x-X)\right)}{\left(5+\sinh^2\left(\sqrt{5}X\right)\right)^2}=\ln\left(\frac{5+2\sqrt{5}}{5-2\sqrt{5}}\right)+4\sqrt{5}$.

The above equation shows that the peak of the soliton energy moves under the influence of a complicated force which is a function of soliton position, it's velocity and also characters of the external potential, $V(x)$.   Equation (\ref{10}) clearly shows that the soliton mass is a space dependent function which is an interesting non-classical behaviour. The soliton energy in the presence of the potential $V(x)=\varepsilon\delta(x)$ is calculated as: 
\begin{equation}\label{11}\nonumber
E=\frac{1}{2}\dot{X}^2\left(M_0-50\varepsilon\frac{\cosh^2\left(\sqrt{5}X\right)}{\left(5+\sinh^2\left(\sqrt{5}X\right)\right)^2}\right)\\
\end{equation}
\begin{equation}
+50\varepsilon\frac{\cosh^2\left(\sqrt{5}X\right)}{\left(5+\sinh^2\left(\sqrt{5}X\right)\right)^2}+M_0.
\end{equation}
It is the energy of a particle with a space dependent mass
\begin{equation}\label{12}
M(X)=M_0-50\varepsilon\frac{\cosh^2\left(\sqrt{5}X\right)}{\left(5+\sinh^2\left(\sqrt{5}X\right)\right)^2}
\end{equation}
and velocity $\dot{X}$ which is moved under the influence of external effective potential. By calculating $\dot{X}^2$ from (\ref{10}) and inserting it into the soliton energy (\ref{11}), one can show that the soliton total energy is a function of the soliton initial position $X_{0}$ and initial velocity $\dot{X}_{0}$. Therefore the energy of a soliton remains conserved during the interaction. Figure \ref{fig1} presents the energy of a static soliton as a function of its position in the potential $V(x) = +0.35\delta(x)$. Because of the extended nature of the soliton, the effective potential is not an exact delta function. 

\begin{figure}
\epsfig{file=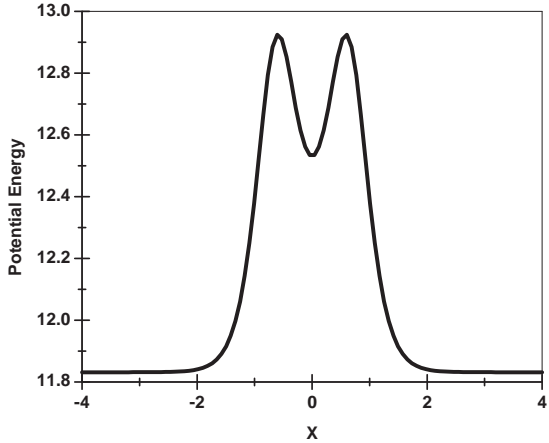,width=\linewidth}
\caption{\label{fig1} The soliton potential energy as a function of collective variable $X$, for $\varepsilon=+0.35$.}
\end{figure}
Figure \ref{fig1} also shows that the energy has two absolute maximum $(E_{max})$ in $X_{m}=\pm\frac{1}{\sqrt{5}}\sinh^{-1}(\sqrt{3})$ and a local minimum $(E_{min})$ in $X=0$. This configuration creates very interesting features for the soliton during the interaction with the potential. There are three different trajectories for a soliton according to its initial conditions and the potential characters. 

\subsection{\label{a}Soliton-barrier system}
Consider a soliton which goes toward the potential barrier $\left( \varepsilon>0 \right)$ from an initial position $\left|X_0\right|>\frac{1}{\sqrt{5}}\sinh^{-1}(\sqrt{3})$ with an initial velocity $\dot{X}_0$. The soliton will reflect back if its total energy is less than the potential maximum $\left( E_{max} \right)$. In this case the soliton initial velocity $\dot{X}_{0}$ is lower than the critical velocity
\begin{equation}\label{13}
v_{c}=\sqrt{\frac{\varepsilon}{M_0-50\varepsilon\frac{\cosh^2\left(\sqrt{5}X_0\right)}{\left(5+\sinh^2\left(\sqrt{5}X_0\right)\right)^2}}}\frac{5\left(3-\sinh^2\left(\sqrt{5}X_0\right)\right)}{2\left(5+\sinh^2\left(\sqrt{5}X_0\right)\right)}.
\end{equation}
 The soliton passes through the barrier if its initial velocity is greater than the critical velocity. In this situation the soliton energy will be greater than the $E_{max}$ . Figure \ref{fig2} demonstrates these two soliton trajectories during the interaction with the potential barrier of $\varepsilon=0.2$ plotted by solving the equation (\ref{10}) numerically. Dashed line presents the trajectory of a soliton with an initial velocity lower than the critical velocity. The soliton climbs the barrier but it can't pass the barrier and reflects back. The solid line shows that a soliton with a velocity greater than the $v_{c}$ passes through the barrier. Note the small fluctuations in the soliton trajectory on top of the barrier. This part of trajectory contains different physics which is discussed with more details in the following. 
 \begin{figure}
\epsfig{file=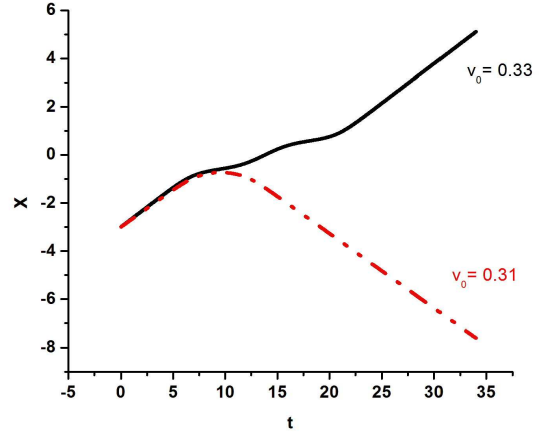,width=\linewidth}
\caption{\label{fig2} Soliton trajectories as function of time with different initial velocity. For the dashed line iinitial velocity has been taken lower than the critical velocity while solid line shows trajectory of a  soliton with initial velocity greater than the critical velocity. Initial position is  $X_0=-3$.}
\end{figure}

Now consider a soliton Which is initially located somewhere in the valley between two peaks of the potential i.e. $\left|X_{0}\right|<\frac{1}{\sqrt{5}}\sinh^{-1}(\sqrt{3})$ with an energy less than the potential maximum, $E<E_{max}$  . The soliton oscillates around the potential minimum which is located at the origin. This situation is unique for a soliton-barrier interaction. It is a very interesting and important behavior. A soliton in this situation can be trapped by a potential barrier which has not been observed before. 

For small amplitude oscillation ( sufficiently small $X_{0}$)  one can use a Taylor expansion for the mass term and the potential energy around $X=0$ in order to find the oscillation frequency as:
\begin{equation}\label{14}
M(X)=M_0-2\varepsilon+\mathrm{O}(X^2)
\end{equation} 
\begin{equation}\label{15}
U(X)=2\varepsilon+6\varepsilon X^2+\mathrm{O}(X^3).
\end{equation} 
With this approximation the angular frequency of the soliton oscillation becomes:
\begin{equation}\label{16}
\omega_{\mathrm{b}}=\sqrt{\frac{12\varepsilon}{M_0-2\varepsilon}}.
\end{equation} 

\subsection{\label{b}Soliton-well system}
A potential well can be created with a negative $\epsilon$, $\epsilon<0$. Therefore the soliton equation of motion becomes:
\begin{equation}\label{17}\nonumber
\ddot{X}\left(M_0+50\varepsilon\frac{\cosh^2\left(\sqrt{5}\left(X\right)\right)}{\left(5+\sinh^2\left(\sqrt{5}(X)\right)\right)^2}\right)=\\
\end{equation}
\begin{equation}
-\left(\dot{X}^2-2\right)50\sqrt{5}\varepsilon\frac{\cosh\left(\sqrt{5}X\right)\sinh\left(\sqrt{5}X\right)\left(3-\sinh^2\left(\sqrt{5}X\right)\right)}{\left(5+\sinh^2\left(\sqrt{5}(x-X)\right)\right)^3}
\end{equation}
and it's energy is
\begin{eqnarray}\label{18}\nonumber
E&=&\frac{1}{2}\dot{X}^2\left(M_0+50\varepsilon\frac{\cosh^2\left(\sqrt{5}\left(X\right)\right)}{\left(5+\sinh^2\left(\sqrt{5}(X)\right)\right)^2}\right)\\
&-&
50\varepsilon\frac{\cosh^2\left(\sqrt{5}\left(X\right)\right)}{\left(5+\sinh^2\left(\sqrt{5}(X)\right)\right)^2}+M_0.
\end{eqnarray}
The equation (\ref{18}) shows that the soliton moves under the influence of an attractive potential. Figure \ref{fig3} shows the potential well with $\varepsilon=-0.35$. 
\begin{figure}
\epsfig{file=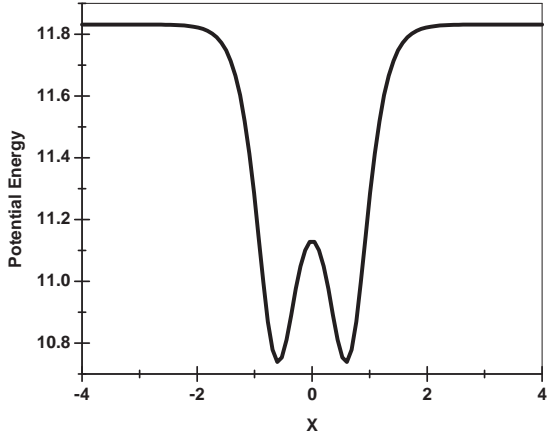,width=\linewidth}
\caption{\label{fig3} The potential energy of soliton as a function of soliton position with $\varepsilon=-0.35$.}
\end{figure}

Assume that a soliton moves toward the centre of the potential well from an initial position $X_0$ with an initial velocity $\dot{X}_0$. The soliton can escape to infinity if its initial velocity is greater than the escape velocity
\begin{equation}\label{19}
v_{\mathrm{escape}}=\frac{\varepsilon}{M_0}\frac{10\cosh\left(\sqrt{5}X_0\right)}{\sqrt{\frac{50\varepsilon}{M_0}\cosh^2\left(\sqrt{5}X_0\right)+\left(5+\sinh^2\left(\sqrt{5}X_0\right)\right)^2}}.
 \end{equation}
But the soliton will captured by the well and oscillates there if its initial speed is lower than the escape velocity $v_{\mathrm{escape}}$. The maximum distance between the soliton and the center of the potential is calculated using (\ref{17}) and (\ref{18}) as
\begin{equation}\label{20}\nonumber
100\varepsilon\frac{\cosh^2\left(\sqrt{5}X_{\mathrm{max}}\right)}{\left(5+\sinh^2\left(\sqrt{5}X_{\mathrm{max}}\right)\right)^2}=
\end{equation}
\begin{equation}
\left(2-\dot{X}_0^2\right)\left(M_0+50\varepsilon\frac{\cosh^2\left(\sqrt{5}X_{0}\right)}{\left(5+\sinh^2\left(\sqrt{5}X_{0}\right)\right)^2}\right)-2M_0.
\end{equation}

The captured soliton has two different oscillation modes according to its total energy. If its energy is greater than $M_{0}-2\varepsilon$ it will oscillate around the center of the well, $X=0$. 
The angular frequency can be calculated using the Taylor series expansion of $M(X)$ and $U(X)$  around $X=0$ as
\begin{equation}\label{21}
M(X)=M_0+2\varepsilon+\mathrm{O}\left(X^2\right)
\end{equation}
and
\begin{equation}\label{22}
U(X)=2\varepsilon+\frac{12\varepsilon}{2} X^2+\mathrm{O}\left(X^3\right).
\end{equation}
Therefore we have
\begin{equation}\label{23}
\omega_{\mathrm{w}}=\sqrt{\frac{12\varepsilon}{M_0+2\varepsilon}}.
\end{equation}

A trapped soliton with an energy less than $M_{0}-2\varepsilon$ oscillates around one of the two degenerate minima of the potential, $X_{min}=\pm\frac{1}{\sqrt{5}}\sinh^{-1}(\sqrt{3})$. The angular  frequency for this oscillation mode is calculated as follows 
\begin{eqnarray}\label{24}\nonumber
M(X)&=&M\left(X=X_1\right)+\left(X-X_1\right)\frac{\partial M}{\partial \mathrm{X}}\big|_\mathrm{X=X_1}+\mathrm{O}\left(\left(X-X_1\right)^2\right) \\
&=&
M_0+\frac{25}{8}\varepsilon+\mathrm{O}\left(\left(X-X_1\right)^2\right)
\end{eqnarray}
and
\begin{equation}\label{25}
U(X)=\frac{25}{2}\varepsilon+\frac{375\varepsilon}{32}(X-X_1)^2+\mathrm{O}\left(\left(X-X_1\right)^3\right)
\end{equation}
and therefore
\begin{equation}\label{26}
\omega_1=\sqrt{\frac{375\varepsilon}{16M_0+50\varepsilon}}.
\end{equation}

\section{\label{third}Numerical simulation}
The dynamics of soliton-potential interaction can be studied using the above results theoretically. Equation \ref{3} also has been solved numerically using the fourth-order Runge-Kutta method for time derivatives. Space derivatives were expanded using the finite difference method. The Hamiltonian density has been calculated using finite difference method in each time step. The delta function has been simulated using Gaussian function $V(x)=\sqrt{\frac{b}{\pi}}\varepsilon e^{-bx^2}$. The stability of numerical procedures and the validity of the results have been checked by doing the calculations with different values for the grid space lengths (0.01, 0.05 and 0.001). Time steps were chosen less than $\frac{1}{4}$ of the grid space steps because of numerical stability considerations.

It is clear that the results of numerical simulations are different from the analytical results because of some used approximations. But one can fit the numerical outcome on the analytical equations using an effective value for the potential strength $\varepsilon$. Thus the numerical results for the critical velocity of a soliton to pass over the potential barrier has been fitted on the derived analytical equation (\ref{13}). The critical velocity can be found numerically by sending a soliton with different initial speed and observing the final situation after the interaction (falling back or getting over the potential). The critical velocity for a soliton which goes toward the potential from infinity is 
\begin{equation}\label{27}
v_c=\frac{5}{2}\sqrt{\frac{\varepsilon}{M_0}}. 
\end{equation}
The soliton initial position $X_{0}=-8$ was chosen in the numerical simulations and the critical velocity has been found for different values of the potential strength. An effective strength can be found by fitting the simulation results on the function $\frac{5}{2}\sqrt{\frac{\varepsilon_\mathrm{{eff}}}{M_0}}$ where $\varepsilon_{\mathrm{eff}}=p_{1}+p_{2}\varepsilon$. Figure \ref{fig4} presents the critical velocity (\ref{27}) and Corresponding numerical results. The effective potential is fitted on the analytic model by $p_1=0.02813\pm 0.00167$ and $p_2=0.56289\pm 0.00556$ with standard deviation of $1.38\times 10^{-6}$ which means $\varepsilon_{\mathrm{eff}}\approx \frac{\varepsilon}{2}$. As mentioned before, the effective potential in the curved space-time is two times of the corresponding potential in the flat space-time. This clearly shows that the theoretical model is described the real situation with a very good approximation.        \begin{figure}
\epsfig{file=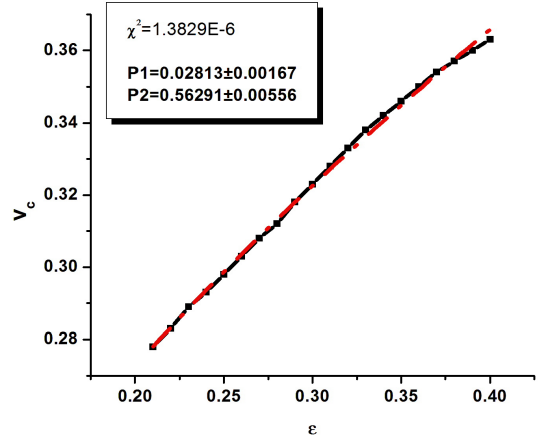,width=\linewidth}
\caption{\label{fig4} Critical velocity as a function of $\varepsilon$. Solid line presents the numerical results while the dashed line is fitted line on the numerical outcome.}
\end{figure}

Equation (\ref{13}) shows that the soliton critical velocity is a function of its initial position. It is expected that one can successfully fit the numerical results on this equation using the calculated $\varepsilon_{\mathrm{eff}}$. Figure \ref{fig5} presents the critical velocity as a function of soliton initial position. The dashed line is plotted using the equation (\ref{13}) and the solid line shows the numerical results. This figure demonstrates a good agreement between the analytical and numerical results.      
\begin{figure}
\epsfig{file=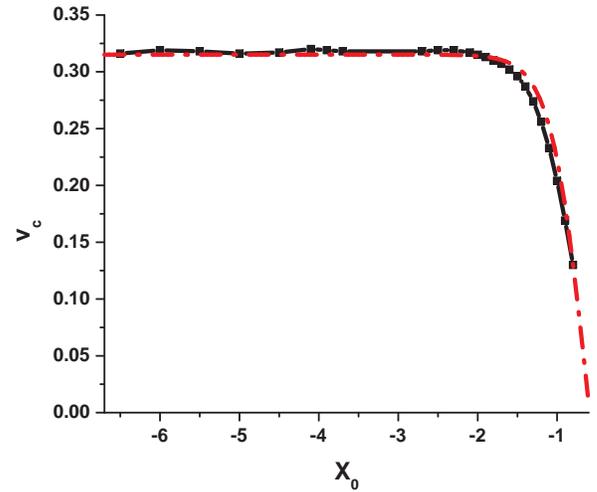,width=\linewidth}
\caption{\label{fig5} Critical velocity as a function of solitin initial position, $X_0$. Solid line shows the numerical results and dashed line is corresponding analytical prediction.}
\end{figure}

Unfortunately such this agreement between the numerical and analytical results have not seen in some of other predictions. Figure \ref{fig6} demonstrates the period of oscillation in a soliton-barrier system. The dashed line has been plotted using the equation (\ref{16}) while the solid line shows the numerical results with corresponding $\varepsilon_{\mathrm{eff}}$. Both curves shows that the oscillation period decreases with an increasing $\varepsilon$. But the predicted period from the analytical model is very different with numerical results. The situation is better for small oscillations in soliton-well system as figure \ref{fig8} presents. The results of numerical simulations has been shown with the solid line, while the dash line has been plotted using equation (\ref{26}).   
\begin{figure}
\epsfig{file=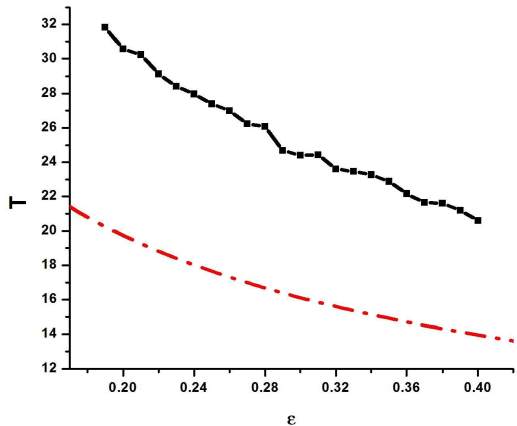,width=\linewidth}
\caption{\label{fig6} The soliton oscillation period (Captured in the center of the barrier) as a function of $\varepsilon$. The solid line shows numerical results and the dashed line has been plotted using (\ref{13}).}
\end{figure}
\begin{figure}
\epsfig{file=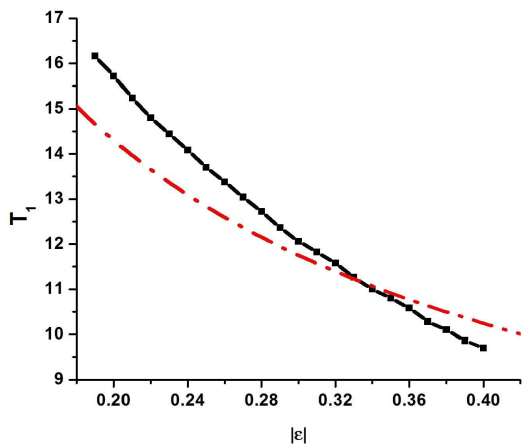,width=\linewidth}
\caption{\label{fig8} Period of oscillation around one of the two degenerate minima of the potential well as a function of $\varepsilon$. Dashed line presents the analytical prediction and solid line shows the numerical results.}
\end{figure}

Interaction of DSG solitons with external defect also has been studied in \cite{40} using collective coordinate approach calculated with different method for adding the potential to the equation of motion. Comparing the predictions with related numerical simulations has not been done in this paper. The general behavior of the soliton predicted in \cite{40} is the same as what has been presented here. This means that two models describe the general features of the soliton-potential interaction correctly. But there are some important differences between two models in the details of interactions. It is because of the different natures of the ways for adding the potential. In our model, solitons have dynamical space dependent mass while the soliton mass in the model \cite{40} is constant. Our presented model saves the symmetries and also topological properties of the theory in the presence of the external potential, but the model in \cite{40} is not able to carry all of the field properties correctly. Note that the solitonic solution in the sine-Gordon model essentially establishes because of topological boundary conditions.       

\section{\label{forth}Conclusions and Remarks}
An analytical model for soliton-potential interaction is presented in double sine-Gordon field theory. Most of the soliton characters during the interaction with potential walls and also potential wells have been derived theoretically. The critical velocity of the soliton to pass over the potential barrier is derived as a function of soliton initial conditions and potential characters. The model predicts that the soliton of double sine-Gordon model can be trapped by a potential barrier which is a very interesting situation. Outcomes of this behavior in applied physics are very important. The period of small amplitude oscillations in this situation has been calculated theoretically. In a potential well, a soliton needs a velocity greater than an escape velocity (or a minimum energy) to be able to go to infinity. The escape velocity has been calculated using the presented model. Two different modes of soliton oscillation in the potential well also have been calculated. 

The analytical results have been compared with the results of direct numerical simulation of the soliton-potential interaction too. In most of the cases analytical and numerical results are in agreement with each other but there are meaningfull differences between the oscillation periods which are derived from analytical model and what has been calculated by numerical simulations. These differences need more attentions. It is possible that the differences come from the interaction between the soliton internal modes with the potential. It is expected that one can resolve this point with an improved model containing better collected coordinate systems. This approach also can be used to create suitable analytical model in other field theories.

\end{document}